\documentclass[12pt]{article}
\usepackage{cite}

\def\be{\begin{equation}}
\def\ee{\end{equation}}
\def\bea{\begin{eqnarray}}
\def\eea{\end{eqnarray}}

\def\tn{\phi_n}

\def\tm{\theta_m}

\usepackage{amsmath}
\usepackage{amssymb}
\usepackage{epsfig}

\newtheorem{theorem}{Theorem}[section]
\newtheorem{proposition}{Proposition}[section]

\begin{document}


\begin{titlepage}

\renewcommand{\thefootnote}{\alph{footnote}}
\vspace*{-3.cm}
\begin{flushright}

\end{flushright}

\vspace*{0.3in}

{\begin{center} {\Large\bf Trigonometrical sums connected with  the chiral Potts model, Verlinde dimension formula,  two-dimensional resistor network, and number theory }

\end{center}}

\vspace*{.8cm} {\begin{center} {\large{\sc
               
                }}
\end{center}}
\vspace*{0cm} {\it
\begin{center}
 \vspace*{.8cm} {\begin{center} {\large{\sc
                Noureddine~Chair
                }}
\end{center}}
\vspace*{0cm} {\it
\begin{center}
 Physics Department,
The  University of Jordan, Amman, Jordan

\end{center}}
\end{center}}
\begin{center}

     Email: n.chair@ju.edu.jo\\ \hspace{19mm}\\
\end{center}

\vspace*{1.5cm}

\begin{center} Abstract\end{center} 

\ \  We have recently  developed   methods for obtaining exact two-point resistance of the complete graph minus $N$ edges. We use these  methods  to obtain closed formulas of certain trigonometrical sums that arise in connection with one-dimensional lattice, in proving the Scott's conjecture on permanent of Cauchy matrix, and in  the perturbative  chiral Potts model. The generalized trigonometrical sums of the chiral Potts model are shown to satisfy recursion formulas that are transparent and direct, and differ from  those of Gervois and  Mehta. By making a change of variables in these recursion formulas, the dimension of the space of conformal blocks of $SU(2)$ and $SO(3)$ WZW models  may be computed recursively.  Our methods are then  extended to compute the corner-to-corner resistance, and the Kirchhoff index  of the first non-trivial two-dimensional resistor network, $2\times N$. Finally,  we  obtain new closed formulas for variant of trigonometrical sums, some of which  appear in connection with number theory.
 \vspace*{.5cm}
\end{titlepage}

\renewcommand{\thefootnote}{\arabic{footnote}}
\section{Introduction}
\ \ We have recently developed  methods for obtaining exact two-point resistance of certain circulant graph namely, the complete graph minus $N$ edges \cite {Chair1}. In this paper, using  similar techniques and ideas, we    consider  trigonometrical sums that arise in the computation of the two-point resistance of the finite  resistor networks \cite{Wu}, in  the work of  McCoy  and Orrick  on the chiral Potts model \cite {McCoy}, and in the Verlinde dimension formula of the  twisted /untwisted space of  conformal blocks of the $SO(3)$/$SU(2)$ WZW model \cite{Verlinde}.

\ \  Before considering these trigonometrical sums, we test the techniques  used in \cite{Chair1}, by first deriving the Green's function of the one-dimensional lattice graphs with free boundaries, and the two-point resistance of the $N$-cycle graph \cite{Wu}. The same techniques is then used to  evaluate  a trigonometrical sum that played a crucial role to prove  R. F Scott's  conjecture  on the permanent of the Cauchy matrix \cite{Minc,Todd}. Having tested these techniques, an  alternative derivation is then given for certain trigonometrical sum that appeared in the perturbative  chiral Potts model \cite {McCoy}, \cite {Mehta}.

\ \  We have also considered the general case studied by  Gervois and Mehta \cite {Mehta},  Berndt and Yeap \cite {Berndt}, here, our results agree with those in \cite {Mehta}.  It turns out that the Verlinde's dimension formulas for the untwisted space of conformal blocks, may be obtained simply by summing over  certain parameter of a trigonometrical sum  considered in \cite{Mehta}.  For the twisted space of conformal blocks, however,  the parameter is restricted to take some value. It is shown that the dimension of the conformal blocks on a genus $g\geq 2$ Riemann surface may be obtained  through a recursion formula that relates  different genera. Mathematically speaking, the dimension of the space of conformal blocks is obtained by expanding  certain generating function order by order, or using the Hirzebruch-Rlemann-Roch theorem \cite{Zagier}.

\ \   By using the method given in \cite {Chair1}, we are able to obtain closed form formula for the  two-point resistance of a $2\times N $ resistor network \cite {Chair2}. In this paper, an exact computation of the corner-to-corner resistance as well as the total effective resistance of a $2\times N$ will be given. The total effective effective resistance, also called the Kirchhoff index \cite{Randic}, this  is an invariant quantity of the resistor network or graph.

\ \   The exact  two-point resistance of an $M\times N$ resistor network is given in terms of a double sum and not in a  closed form \cite {Wu}. Therefore, our computation carried out in this paper, represents the first non-trivial exact results for the two-point resistance of a two-dimensional resistor network.

\ \ This method is then used  to evaluate variant of trigonometrical sums, some of which are related to number theory, we hope that these trigonometrical sums will have some physical applications.  It is  interesting to point out that  all the computations of the trigonometrical sums  in this  paper  are based on a  formula by Schwatt \cite{ Schwatt} on trigonometrical power sums, and  the representation  of the binomial coefficients  by the  residue operator. The  Schwatt's formula is modified slightly,  only in the case of the corner-to-corner resistance, the Kirchhoff index  and  trigonometrical sum given by $ F_{1}(N,l,2) $, and $ F_{1}(N,l,2)$, see Section 6, this was also the case in our previous paper  \cite{Chair1}.

\ \ This paper is organized as follows; in section 2, we give an explicit computations of the two-point resistance of the $N$-cycle graph and the Green's function of the one-dimensional lattice, and in Section 3,  we give a simple  derivation of a trigonometrical sum connected with Scott's conjecture on the permanent of the Cauchy matrix. In section 4, we consider trigonometrical sums  arising in the chiral Potts model, and in the Verlinde formula of the dimension of the conformal blocks. Exact computations of the corner-to-corner and the Kirchhoff index of $2\times N$ resistor network will be given  in section 5. In Section 6, we consider other class of trigonometrical sums  some of which are related to number theory, and  finally, in Section 7,  our conclusions are given.
\section{ The two-point resistance of one-dimensional lattice using the residue operator }

\ \ In this Section, we first  start with the two-point resistance of the $N$-cycle graph computations, then, move to the trigonometrical sum related to the two-point resistance of the one-dimensional lattice with free boundaries, that is, the  path graph. The two-point resistance of the $N$-cycle graph  between any  two nodes $\alpha$ and $\beta$ is  given by the following simple closed formula \cite{Wu},
\begin{equation} 
\label{t1}
R(l)=\frac{1}{N}\sum_{n=1}^{N-1}\frac{\sin^2(nl\pi/N)}{\sin^2(n\pi/N)}=\frac{l(N-l)}{N},
\end{equation}
where $l=|\alpha-\beta|$, and $1\leq \alpha,\beta \leq N. $
Our derivation for the two-point resistance starts with the following trigonometrical identity  $$\cos(2ln\pi/N)=\sum_{s=0}^{l}(-1)^s \frac{l}{l+s}\binom {l+s} {l-s}2^{2s}\sin^{2s}(n\pi/N), $$ from which the  above trigonometrical sum may be rewritten as
\begin{equation} 
\label{t2}
R(l)=\frac{1}{2N}\sum_{s=1}^{l}(-1)^{s +1}\frac{l}{l+s}\binom {l+s} {l-s}2^{2s}\sum_{n=1}^{N-1} \sin^{2(s-1)}(n\pi/N).
\end{equation}  
On the other hand, Schwatt's formula for trigonometrical power sums \cite{Schwatt}, gives
\begin{equation} 
\label{t3}
\sum_{n=1}^{N-1} \sin^{2(s-1)}(n\pi/N)=\frac{1}{2^{2(s-1)-1}}\sum_{t=1}^{s-1}(-1)^{t +1}\binom {2(s-1)}{s-1-t} +\frac{N-1}{2^{2(s-1)}}\binom {2(s-1)} {s-1}.
\end{equation}

\ \ 
 Therefore,  the two-point resistance may be obtained by evaluating  the binomial sums in the expression of $R(l)$, based on the  residue operator. This operator played a crucial  role in evaluating combinatorial sums and proving combinatorial identities \cite{Egorychev}.  First, let us recall the definition of the residue operator $\hbox{res}$. To that end, let  $G(w)= \sum_{k=0}^{\infty}a_{k}w^{k}$ be a generating function for a sequence $\{a_{k}\}$. Then the k-th coefficient of $G(w)$ may be represented by the formal residue as follows
$$a_{k}= \hbox{res}_w G(w){w^{-k-1}}.$$
This is equivalent  to the Cauchy integral representation of $a_k$,
\[a_k=\frac{1}{2\pi i}\oint _{|z|=\rho}\frac{G(w)}{w^{k+1}}dw ,\] for coefficients of the Taylor series in a punctured neighborhood of zero. In particular, the generating function of the binomial coefficient sequence  $\binom {n}{k}$ for a fixed $n$ is given by $$G(w)= \sum_{k=0}^{n}\binom {n}{k} w^{k} =(1+w)^{n},$$ 
and hence $$\binom {n}{k}=\hbox{res}_w (1+w){^n}{w^{-k-1}}.$$ The other binomial coefficient that we need in this paper is the following, $$\binom {2n}{n}=\hbox{res}_w (1-4w){^{-1/2}}{w^{-n-1}}.
$$

\ \  Before finishing this brief summary, we should mention  one important property of the  residue operator $\hbox{res}$, namely linearity, this is a  crucial in doing  computations,  linearity states that;  given some contants $\alpha$ and $\beta$, then $$\alpha \hbox{res}_w G_{1}(w){w^{-k-1}}+ \beta\hbox{res}_w G_{2}(w){w^{-k-1}}=\hbox{res}_w(\alpha G_{1}(w)+ \beta G_{2}(w)) {w^{-k-1}}.$$ 
Let us now evaluate the first term in Eq.(\ref{t2}), using the residue operator, namely the following term
\begin{eqnarray}
\label{t4}
R_{1}(l):&=&\frac{2}{N}\sum_{s=1}^{l}(-1)^{s }\frac{2l}{l+s}\binom {l+s} {l-s}\sum_{t=1}^{s-1}(-1)^{t }\binom {2(s-1)}{s-1-t}\nonumber\\&=&\frac{2}{N}\sum_{s=1}^{l}(-1)^{s }\frac{2l}{l+s}\binom {l+s} {l-s}\sum_{t=1}^{s-1}(-1)^{t }\hbox{res} \frac{(1+w)^{2(s-1)}}{w^{s-t}}\nonumber\\&=&\frac{2}{N}\sum_{s=1}^{l}(-1)^{s }\frac{2l}{l+s}\binom {l+s} {l-s}\hbox{res}_{w=0}\frac{(1+w)^{2(s)}}{(1+w)^{3}w^{s-1}}.
\end{eqnarray}
In obtaining Eq.(\ref{t4}), we  discarded an analytic term at the pole $w=0$ of order $s-1$. By making a change of variable $l-s=k $, then, Eq. (\ref{t4}) may be rewritten as 
\begin{eqnarray}
\label{t5}
R_{1}(l)&=&(-1)^{l+1}\frac{2}{N}\hbox{res}_{w=0}\frac{w}{(1+w)^{3}}\sum_{k=1}^{l-1}(-1)^{ k}\frac{2l}{2l-k}\binom {2l-k} {k}\bigg(\frac{1+w}{\sqrt{w}}\bigg)^{2(l-k)}\nonumber\\&=& (-1)^{l+1}\frac{2}{N}\hbox{res}_{w=0}\frac{w}{(1+w)^{3}}\bigg(C_{2l}\big(\frac{1+w}{\sqrt{w}}\big)-(-1)^{l}\bigg),
\end{eqnarray}
where $$ C_{2l}(x)= 2T_{2l}(x/2)=\sum_{k=0}^{l}(-1)^{k} \frac{2l}{2l-k}\binom {2l-k} {k}x^{2l-2k},$$
is the normalized Chebyshev polynomial of the first kind\cite{Rivlin}, and
$$T_{2l}(x/2)=\frac{1}{2}\Bigg\lbrack\Bigg(\frac{x}{2}+\sqrt{(x/2)^2-1}\bigg)^{2l} +\Bigg(\frac{x}{2}-\sqrt{(x/2)^2-1}\bigg)^{2l} \Bigg\rbrack.$$
Using the fact that $C_{2l}\big(\frac{1+w}{\sqrt{w}}\big)= \frac{1}{w^{l}}+w^{l}$, then the final expression for the first term $R_{1}(l) $, reads
\begin{eqnarray}
\label{t6}
R_{1}(l)&=&(-1)^{l+1}\frac{2}{N}\hbox{res}_{w=0}\frac{1}{(1+w)^{3}w^{l-1}}\nonumber\\&=&-\frac{l(l-1)}{N}
\end{eqnarray}
Similarly, the second term may written as 
\begin{eqnarray}
\label{t7}
R_{2}(l):&=&\frac{(N-1)}{N}\sum_{s=1}^{l}(-1)^{s }\frac{2l}{l+s}\binom {l+s} {l-s}\binom {2(s-1)} {s-1}\nonumber\\&=&(-1)^{l+1}\frac{(N-1)}{N}\hbox{res}_{w=0}\frac{1}{(1+w)^{2}w^{l}}\nonumber\\&=& \frac{l(N-1)}{N}.
\end{eqnarray}
 Adding the contributions given by Eqs. (\ref{t6}), and  (\ref{t7}), then, we get  
\begin{equation}
\label{t8}
R(l)=\frac{1}{N}\sum_{n=1}^{N-1}\frac{\sin^2(nl\pi/N)}{\sin^2(n\pi/N)}=\frac{l(N-l)}{N}.
\end{equation}

\ \   Now, we want to evaluate the following trigonometrical sum $$  F_N(l ) = \frac {1} {N}\sum_{n=1}^{N-1} \frac {1-\cos nl\pi/N} {1-\cos n\pi/N},$$ this sum arises in connection with the two-point resistance of a path graph \cite{Wu}. The evaluation of this term may be done as follows; 
\begin{eqnarray}
\label{t9}
F_N(l ) &= &\frac {1} {N}\sum_{n=1}^{N-1} \frac {1-\cos nl\pi/N} {1-\cos n\pi/N}\nonumber\\&=&\frac {1} {N}\sum_{n=1}^\frac{N}{2} \frac {1-\cos(2n-1)l\pi/N} {1-\cos(2n-1)\pi/N}+\frac {1} {N}\sum_{n=1}^{\frac{N}{2}-1} \frac {1-\cos 2nl\pi/N} {1-\cos2n\pi/N},
\end{eqnarray}
here,  $N$ is  assumed to be even, similar steps may be used for  $N$ odd. It is interesting to note that in evaluating $F_N(l )$, we only need to compute the first term since the second term is related to the two-point resistance of the $N$-cycle graph given by Eq. (\ref{t8}). Then, the first term may be written as 
\begin{eqnarray}
\label{t10}
\frac {1} {N}\sum_{n=1}^\frac{N}{2} \frac {1-\cos(2n-1)l\pi/N} {1-\cos(2n-1)l\pi/N}&=&\frac {1} {N} \sum_{n=1}^\frac{N}{2} \frac {\sin^{2}(2n-1)l\pi/2N} {\sin^{2}(2n-1)l\pi/2N}\nonumber\\&=&\frac{1}{2N}\sum_{s=1}^{l}(-1)^{s +1}\frac{l}{l+s}\binom {l+s} {l-s}2^{2s}\sum_{n=1}^\frac{N}{2}\sin^{2(s-1)}\frac{(2n-1)\pi}{2N}.\nonumber\\.
 \end{eqnarray}
 By using the identity $$\sum_{n=1}^\frac{N}{2}\sin^{2(s-1)}(2n-1)\pi/2N= \frac{1}{2}\bigg(\sum_{n=1}^{N-1}\sin^{2(s-1)}n\pi/2N+\sum_{n=1}^{N-1}(-1)^{n-1}\sin^{2(s-1)}n\pi/2N\bigg),  $$ 
 and the formulas for trigonometrical power sums given in \cite{Schwatt}, then, one can show
\begin{eqnarray}
\label{t11}
\sum_{n=1}^\frac{N}{2}\sin^{2(s-1)}(2n-1)\pi/2N=\frac{2N}{2^{2s}}\binom {2(s-1)} {s-1},
\end{eqnarray}
which in turn, implies that the formula for the first term  should be
 \begin{eqnarray}
\label{t12}
\frac {1} {N}\sum_{n=1}^\frac{N}{2} \frac {1-\cos(2n-1)l\pi/N} {1-\cos(2n-1)\pi/N}&=&\frac{1}{2}\sum_{s=1}^{l}(-1)^{s +1}\frac{2l}{l+s}\binom {l+s} {l-s}\binom {2(s-1)} {s-1}\nonumber\\&=&\frac{l}{2}.
\end{eqnarray}
To compute the second term given in Eq. (\ref{t9}), we use the following symmetry 
enjoyed by the two-point resistance of the $N$-cycle graph,   $N$ even
\begin{eqnarray}
\label{t13}
\frac {1} {N}\sum_{n=1}^{N-1} \frac {1-\cos 2nl\pi/N} {1-\cos2n\pi/N}&=&\frac {2} {N}\sum_{n=1}^{\frac{N}{2}-1} \frac {1-\cos 2nl\pi/N} {1-\cos2n\pi/N}+\frac {1} {2N}\big(1-(-1)^{l}\big).
\end{eqnarray}
Therefore, the second term may be obtained to give the following closed formula for $F_N(l ) $
\begin{equation}
\label{t14}
F_N(l ) = \frac {1} {N}\sum_{n=1}^{N-1} \frac {1-\cos nl\pi/N} {1-\cos n\pi/N}= l-\frac{1}{N}\bigg(\frac{l^2}{2}+\frac{1}{4}(1-(-1)^{l})\bigg)
\end{equation}
This is in a complete agreement with the formula for the Green's function for the path graph in \cite{Wu}
\section{Trigonometrical sum connected with  Scott's conjecture}

\ \ 
 In  proving  R. F Scott's conjecture on the permanent  of the Cauchy matrix,  Minc in \cite{Minc} needed to evaluate the following trigonometrical sum;
  \begin{equation}
\label{sc0}
 \sum_{n=1}^{N} \frac {\cos (2n-1)l\pi/N} {1-\cos (2n-1)\pi/N}.
 \end{equation}
 He obtained a closed-form formula for this sum using induction,
 and the sum turns out to  be equal to  $\frac{N}{2}(N-2l)$.  A short  time  later, Stembridge and  Todd \cite{Todd}, gave a proof for the evaluation for this  sum, based on linear algebra. Here, we give a short derivation for this sum using  our formula given by Eq. (\ref{t12}), and the well-known identity 
 \begin{equation}
\label{sc1}
 \sum_{n=1}^{N-1}\frac{1}{\sin^2(n\pi/N)}=\frac{N^{2}-1}{3}.
\end{equation}
Our derivation follows easily by realizing that  Eq. (\ref{t12}) is symmetric under the shift $n\rightarrow N-n$, and as a consequence one  gets 
\begin{equation}
\label{sc2}
\sum_{n=1}^{N} \frac {1-\cos(2n-1)l\pi/N} {1-\cos(2n-1)\pi/N}=N l.
\end{equation}
In order to evaluate the sum in  Eq. (\ref{sc0}), we need a formula for the sum $$\sum_{n=1}^{N} \frac {1} {1-\cos(2n-1)\pi/N} =\frac{1}{2}\sum_{n=1}^{N} \frac {1} {\sin^{2}(2n-1)\pi/2N }.$$
The latter may be evaluated as follows
\begin{eqnarray}
\label{sc3}
\frac{1}{2}\sum_{n=1}^{N} \frac {1} {\sin^{2}(2n-1)\pi/2N}&=&\frac{1}{2}\sum_{n=1}^{2N-1}\frac{1}{\sin^2(n\pi/2N)}-\frac{1}{2}\sum_{n=1}^{N-1}\frac{1}{\sin^2(2n\pi/2N)}\nonumber\\&=&\frac{N^{2}}{2}.
\end{eqnarray}
In obtainnig Eq. (\ref{sc3}), we used the identity given in Eq. (\ref{sc1}), thus, using Eq. (\ref{sc2}) and  Eq. (\ref{sc3}), we may write
\begin{equation}
\label{sc4}
 \sum_{n=1}^{N} \frac {\cos (2n-1)l\pi/N} {1-\cos (2n-1)\pi/N}=\frac{N^2}{2}-Nl.
\end{equation}
This is exactly the result obtained by Minc, Stembridge and  Todd \cite{Minc,Todd}.
\section{Trigonometrical sums arising in the chiral Potts model and in the Verlinde's formula}

\ \ In this section  the  trigonometrical sum $T_{4}(l):=\sum_{n=1}^{N-1}\frac{\sin^2(nl\pi/N)}{\sin^4(n\pi/N)} $ is evaluated in a closed form using the residue operator. We also give an almost closed formula for the general case $T_{2m}(l):=\sum_{n=1}^{N-1}\frac{\sin^{2}(nl\pi/N)}{\sin^{2m}(n\pi/N)} $, for $m\geq1$. The first  trigonometrical sum arises  in the work of  McCoy  and Orrick  on the chiral Potts model \cite {McCoy},  this sum including other trigonometrical identities  were proved  by Gervois and Mehta \cite {Mehta}. The second sum namely the sum  $T_{2m}(l)$,  was considered by  Gervois and Mehta \cite {Mehta} using a recursion formula. Here, we will obtain   recursion formulas for both  $T_{2m}(l)$ and $$T_{2m}:=\sum_{n=1}^{N-1}\frac{1}{\sin^{2m}(n\pi/N)}.$$

\ \  If, we set $N=k+2$ and $m =g-1$, $k,g$ being the level of the $su(2)$ Kac-Moody algebra, and the genus of the Riemann surface respectively. Then, the  sum $ T_{2m}$ up to to some normalization factor  is nothing but the dimension of the space of the conformal blocks of the $SU(2)$ WZW model. As a consequence, the recursion formula derived for  $ T_{2m}$, may be used to obtain the expression for the dimension of the space of  the conformal blocks for a given genus $g$. Similar computations are carried out for the twisted trigonometrical sum $$T_{2m}^{t}:=\sum_{n=1}^{N-1}(-1)^{n+1}\frac{1}{\sin^{2m}(n\pi/N)}.$$ This is related to the dimension of the space of the conformal blocks of the $SO(3)$ WZW model.
 \subsection {Trigonometrical sums and the perturbative chiral Potts model}
 
 \ \  Let us first start with the trigonometrical sums arising in the perturbative treatment  of the the chiral Potts model \cite {McCoy}. Techniques of the previous section, may be used to evaluate the  sum  $T_{4}(l)$, as follows
\begin{eqnarray}
\label{t15}
T_{4}(l)&=&\sum_{n=1}^{N-1}\frac{\sin^2(nl\pi/N)}{\sin^4(n\pi/N)}\nonumber\\&=&l^{2}\sum_{n=1}^{N-1}\frac{1}{\sin^2(n\pi/N)}+\frac{1}{2}\sum_{s=2}^{l}(-1)^{s +1}\frac{l}{l+s}\binom {l+s} {l-s}2^{2s}\sum_{n=1}^{N-1} \sin^{2(s-2)}(n\pi/N)\nonumber\\&=&\frac{l^2}{3}(N^{2}-1)+8\sum_{s=2}^{l}(-1)^{s }\frac{2l}{l+s}\binom {l+s} {l-s}\sum_{t=1}^{s-2}(-1)^{t }\binom {2(s-2)}{s-2-t}\nonumber\\&+& 4(N-1)\sum_{s=2}^{l}(-1)^{s +1}\frac{2l}{l+s}\binom {l+s} {l-s}\binom {2(s-2)}{s-2},
\end{eqnarray}
the first term in the above equation follows from the well-known identity $$\sum_{n=1}^{N-1}\frac{1}{\sin^2(n\pi/N)}=\frac{N^{2}-1}{3},$$ while the second and the third terms may computed using the residue operator as in the previous section to give,
\begin{eqnarray}
\label{t16}
\sum_{s=2}^{l}(-1)^{s +1}\frac{2l}{l+s}\binom {l+s} {l-s}\sum_{t=1}^{s-2}(-1)^{t }\binom {2(s-2)}{s-2-t}&=&(-1)^{l+1} \hbox{res}_{w=0}\frac{1}{(1+w)^{5}w^{l-2}}\nonumber\\&=&\frac{1}{4!}(l+1)l(l-1)(l-2),
\end{eqnarray}
and
\begin{eqnarray}
\label{t17}
\sum_{s=2}^{l}(-1)^{s +1}\frac{2l}{l+s}\binom {l+s} {l-s}\binom {2(s-2)}{s-2}&=& (-1)^{l+1}\hbox{res}_{w=0}\frac{1}{(1+w)^{4}w^{l-1}}\nonumber\\&=&-\frac{1}{3!}(l+1)l(l-1).
\end{eqnarray}
Therefore, the closed formula for the sum given in Eq. (\ref{t15}), reads
\begin{eqnarray}
\label{t18}
\sum_{n=1}^{N-1}\frac{\sin^2(nl\pi/N)}{\sin^4(n\pi/N)}&=&\frac{l^2}{3}(N^{2}-1)+\frac{1}{3}(l+1)l(l-1)(l-2)-\frac{2(N-1)}{3}(l+1)l(l-1)\nonumber\\&=&\frac{l^2}{3}(N-l)(N-l)+\frac{2}{3}(N-l).
\end{eqnarray}
This is exactly the result obtained by  Gervois and Mehta using a recursion formula satisfied by  $T_{2m}(l)$ \cite {Mehta}. Next, we will  give another recursion formula for the  sum  $T_{2m}(l)$. Now, $T_{2m}(l)$, may be written as 
\begin{eqnarray}
\label{t19}
T_{2m}(l)&=&\sum_{n=1}^{N-1}\frac{\sin^2(nl\pi/N)}{\sin^{2m}(n\pi/N)}\nonumber\\&=&\frac{1}{2}\sum_{n=1}^{N-1}\sum_{s=1}^{m-1}(-1)^{s+1 }\frac{l}{l+s}\binom {l+s} {l-s}2^{2s}\frac{1}{\sin^{2(m-s)}(n\pi/N)}\nonumber\\&+&\frac{1}{2}\sum_{s=m}^{l}(-1)^{s +1}\frac{l}{l+s}\binom {l+s} {l-s}2^{2s}\sum_{n=1}^{N-1} \sin^{2(s-m)}(n\pi/N).  
\end{eqnarray}
The first term on the right-hand side, is written in terms of the sum $$T_{2k}=:\sum_{n=1}^{N-1}\frac{1}{\sin^{2k}(n\pi/N)}. $$ This may be  computed  \cite {Mehta}, using $$T_{2k}=\sum_{n=1}^{N-1}\Big(\cot^{2}(\frac{n\pi}{N})+1\Big)^{k}=\sum_{l=1}^{k} \binom{k}{l}S_{l}, $$ where $S_{l}=\sum_{n=1}^{N-1}\Big(\cot^{2}(\frac{n\pi}{N}) \Big)^{2l}$  and a recurrence  relation satisfied by the power sums $S_{l} $. It turns out that  $T_{2k}$,   may also be obtained  using a recursion formula, this will be shown shortly. Now, the second term may be written as follows 
\begin{eqnarray}
\label{t20}
\tilde{T}_{2m}(l):&=&\frac{1}{2}\sum_{s=m}^{l}(-1)^{s +1}\frac{l}{l+s}\binom {l+s} {l-s}2^{2s}\sum_{n=1}^{N-1} \sin^{2(s-m)}(n\pi/N)\nonumber\\&=&2^{2m-1}\sum_{s=m}^{l}(-1)^{s }\frac{2l}{l+s}\binom {l+s} {l-s}\sum_{t=1}^{s-m}(-1)^{t }\binom {2(s-m)}{s-m-t}\nonumber\\&+&2^{2m-2}(N-1)\sum_{s=m}^{l}(-1)^{s +1}\frac{2l}{l+s}\binom {l+s} {l-s}\binom {2(s-m)}{s-m}\nonumber\\&=&(-1)^{l+1}2^{2m-1}\hbox{res}_{w=0}\frac{1}{(1+w)^{2m+1}w^{l-m}}\nonumber\\&+&(-1)^{l+1}2^{2m-2}(N-1)\hbox{res}_{w=0}\frac{1}{(1+w)^{2m}w^{l+1-m}}\nonumber\\&=&(-1)^{m}2^{2m-1}\frac{(l+m-1)!}{(l-m-1)!(2m)!}+ (-1)^{m+1}(N-1)2^{2m-2}\frac{(l+m-1)!}{(l-m)!(2m-1)!}\nonumber\\.
\end{eqnarray}
Therefore, we succeeded  in writing $\tilde{T}_{2m}(l)$  in a closed form  formula,   One can check easily that our results agree with those given in  \cite {Mehta}, and so the formula for  $T_{2m}(l)$ becomes
\begin{eqnarray}
\label{t21}
T_{2m}(l)&=&\sum_{n=1}^{N-1}\frac{\sin^2(nl\pi/N)}{\sin^{2m}(n\pi/N)}\nonumber\\&=&\frac{1}{2}\sum_{s=1}^{m-1}(-1)^{s+1 }\frac{l}{l+s}\binom {l+s} {l-s}2^{2s}T_{2(m-s)}\nonumber\\&+&(-1)^{m+1}2^{2m-1}\frac{(l+m-1)!}{(l-m)!(2m)!}(mN-l).
\end{eqnarray}
setting   $m=1,2$, then, our previous  results given by Eq's. (\ref{t8}) and (\ref{t18}) respectively  are recovered. From Eq. (\ref{t21}), it is clear that  in order to have a closed formula for  $T_{2m}(l)$, one  needs also,  the exact  expressions  for $T_{2k}$, $ k=1\cdots, m-1$. Next, we will  show that $ T_{2k}$ satisfies a recursion formula that involves  the $T_{2k}$'s.  The expression for $T_{2m}$  may be obtained from  $T_{2m}(l)$ from the following simple formula;
\begin{eqnarray}
\label{t22}
\sum_{l=1}^{N-1}T_{2m}(l)&=&\sum_{l=1}^{N-1}\sum_{n=1}^{N-1}\frac{\sin^2(nl\pi/N)}{\sin^{2m}(n\pi/N)}\nonumber\\&=&\frac{N}{2}\sum_{n=1}^{N-1}\frac{1}{\sin^{2m}(n\pi/N)},
\end{eqnarray}
Therefore, we may write
\begin{eqnarray}
\label{t23}
T_{2m}&=&\sum_{n=1}^{N-1}\frac{1}{\sin^{2m}(n\pi/N)}\nonumber\\&=&\frac{1}{N}\sum_{l=1}^{N-1}\sum_{s=1}^{m-1}(-1)^{s+1 }\frac{l}{l+s}\binom {l+s} {l-s}2^{2s}T_{2(m-s)}\nonumber\\&+&\frac{2}{N}\sum_{l=1}^{N-1}(-1)^{l+1}2^{2m-1}\hbox{res}_{w=0}\frac{1}{(1+w)^{2m+1}w^{l-m}}\nonumber\\&+&\frac{2}{N}\sum_{l=1}^{N-1}(-1)^{l+1}2^{2m-2}(N-1)\hbox{res}_{w=0}\frac{1}{(1+w)^{2m}w^{l+1-m}}\nonumber\\&=&\frac{1}{N}\sum_{l=1}^{N-1}\sum_{s=1}^{m-1}(-1)^{s+1 }\frac{l}{l+s}\binom {l+s} {l-s}2^{2s}T_{2(m-s)}\nonumber\\&+&(-1)^{m+1}2^{2m-1}\frac{(N+m-1)!}{N (N-m-1)!(2m+1)!}(2mN+1-N).
\end{eqnarray}
As a result, from our recursion formula,  the different $T_{2m}$'s may be obtained directly, we do not have to  use  the recurrence relation satisfied by the power sum $S_{l}$ \cite {Mehta}. For $m=1$ the first term in Eq. (\ref{t23}) does not contribute and the second term gives the well known formula $ T_{2}=\frac{N^{2}-1}{3}$. Now, for $m=2$, the first term may be computed to give  $ \frac{{(N^2-1)(N-1)}(2N-1)}{9}$, while the second term gives $ -\frac{{(N^2-1)(N-2)}(3N+1)}{15}$, and hence,   $T_{4}= \frac{{(N^2-1)(N^2+11)}}{45}$ in a full agreement with  \cite {Mehta}, \cite { Berndt}.
\subsection{The Verlinde dimension formula}

\ \ The verlinde dimension formula may be obtained simply by setting $m=g-1$,  $N=k+2$ in the expression of  $T_{2m}$, where $g\ge2$, $ k$ are the genus of the Riemann surface, and the level of the lie algebra $SU(2)$, respectively. Then,   $T_{2(g-1)}$ up to some normalization factor, is  the dimension of the space  of conformal blocks $V_{g}$ of the $SU(2)$ WZW model \cite{Verlinde},
 \begin{eqnarray}
\label{t24}
dimV_{g,k}&=&\Big(\frac{k+2}{2}\Big)^{g-1}\sum_{n=1}^{k+1}\frac{1}{\sin^{2g-2}n\pi/(k+2)}\nonumber\\&=&\Big(\frac{k+2}{2}\Big)^{g-1} \frac{1}{(k+2)}\sum_{l=1}^{k+1}\sum_{s=1}^{g-2}(-1)^{s+1 }\frac{l}{l+s}\binom {l+s} {l-s}2^{2s}T_{2g-2 -2s} \nonumber\\&+&(-1)^{g}2^{2g-3}\Big(\frac{k+2}{2}\Big)^{g-1} \frac{1}{(k+2)}\frac{1}{(2g-1)}\binom{k+g}{2g-2}\big((k+2)(2g-3)+1\big).
\end{eqnarray}
 As a consequence, the dimension of the space of the conformal blocks of the $SU(2)$ WZW model,  may be  computed using our recursion formula for $T_{2k}$. Our formula Eq. (\ref{t24}) may be used to give 
\begin{eqnarray}
\label{t25}
dimV_{2,k}&=&\frac{(k+1)(k+2)(k+3)}{6}\nonumber\\dimV_{3,k}&=&\frac{1}{5}\frac{(k+1)(k+2)(k+3)}{6}\Big[\frac{(k+1)(k+2)(k+3)}{6}+2(k+2)\Big]\nonumber\\dimV_{4,k}&=&\frac{1}{7}.\frac{1}{5}\frac{(k+1)(k+2)(k+3)}{6}\nonumber\\&.&\Big[\frac{(k+1)(k+2)(k+3)}{6}\Big[\frac{2(k+1)(k+2)(k+3)+27(k+2)}{6}\Big]+6(k+2)^{2}\Big].\nonumber\\
\end{eqnarray}
Our first two expressions for dimension of the space of conformal blocks agree with those computed using conformal field theory \footnote{The last term $k+2$ in the expression of  $dimV_{3,k}$  should be corrected  in \cite{Piunikhin} in order to make it a positive integer.} \cite{Piunikhin}. For $g=4$, our formula  for $dimV_{4,k} $ is identical to the formula given by Zagier \cite{Zagier} provided  the shift $k\rightarrow k+2$ is taken. This shift is natural, since Zagier defined the dimension of the space conformal  blocks as $dimV_{g,k-2}$.

\ \  For the WZW model based on $SO(3)$, the level $k$ must be even \cite {Thaddeus}, and the formula for the  dimension of the twisted space of the conformal blocks $V_{g,k}^{t}$,  may be written as 
$$dimV_{g,k}^{t}=\Big(\frac{k+2}{2}\Big)^{g-1}\sum_{n=1}^{k+1}(-1)^{n+1}\frac{1}{\sin^{2g-2}n\pi/(k+2)}.$$ In order to  derive a recursion formula for the  dimension $ dimV_{g,k}^{t}$, we first note that the expression for the twisted version of the  trigonometrical sum $ T_{2m}(l)$ given by Eq. (\ref{t19}) is
\begin{eqnarray}
\label{t26}
T_{2m}^{t}(l)&=&\sum_{n=1}^{N-1}(-1)^{n+1} \frac{\sin^2(nl\pi/N)}{\sin^{2m}(n\pi/N)}\nonumber\\&=&\frac{1}{2}\sum_{s=1}^{m-1}(-1)^{s+1 }\frac{l}{l+s}\binom {l+s} {l-s}2^{2s}T_{2(m-g)}^{t}\nonumber\\&+&\frac{1}{2}\sum_{s=m}^{l}(-1)^{s +1}\frac{l}{l+s}\binom {l+s} {l-s}2^{2s}\sum_{n=1}^{N-1} (-1)^{n+1}\sin^{2(s-m)}(n\pi/N),  
\end{eqnarray}
where $$T_{2m}^{t}=\sum_{n=1}^{N-1}(-1)^{n+1}\frac{1}{\sin^{2m}(n\pi/N)} $$
The trigonometrical sum $$ \sum_{n=1}^{N-1} (-1)^{n+1}\sin^{2(s-m)}(n\pi/N),$$ is non-vanishing only if $ N$ is even\cite{Schwatt} and is given by
 \begin{eqnarray}
\label{t27}
 \sum_{n=1}^{N-1} (-1)^{n+1}\sin^{2(s-m)}(n\pi/N)&=&2^{2(m-s)+1}\sum_{t=1}^{s-m}(-1)^{t }\binom {2(s-m)}{s-m-t}\nonumber\\&+&2^{2(m-s)} \binom {2(s-m)}{s-m}
\end{eqnarray}
This formula shows clearly that the dimension of the twisted space of the conformal blocks is is non-vanishing only if $k$ is even, $N=k+2$ in agreement with the algebro-geometrical argument \cite {Thaddeus}. Then, the expression for $T_{2m}^{t}$ may be written as
\begin{eqnarray}
\label{t28}
T_{2m}(l)^{t}&=&\sum_{n=1}^{N-1}(-1)^{n+1}\frac{\sin^2(nl\pi/N)}{\sin^{2m}(n\pi/N)}\nonumber\\&=&\frac{1}{2}\sum_{n=1}^{N-1}(-1)^{n+1}\sum_{s=1}^{m-1}(-1)^{s+1 }\frac{l}{l+s}\binom {l+s} {l-s}2^{2s}\frac{1}{\sin^{2(m-s)}(n\pi/N)}\nonumber\\&+&2^{2m-1}\sum_{s=m}^{l}(-1)^{s +1}\frac{2l}{l+s}\binom {l+s} {l-s}\sum_{t=1}^{s-m}(-1)^{t }\binom {2(s-m)}{s-m-t}\nonumber\\&+&2^{2m-2}\sum_{s=m}^{l}(-1)^{s +1}\frac{2l}{l+s}\binom {l+s} {l-s}\binom {2(s-m)}{s-m}\nonumber\\&=&\frac{1}{2}\sum_{s=1}^{m-1}(-1)^{s+1 }\frac{l}{l+s}\binom {l+s} {l-s}2^{2s}T_{2(m-s)}^{t}\nonumber\\&+&(-1)^{m+1}2^{2m-1}\frac{(l+m-1)!}{(l-m)!(2m)!}l.  
\end{eqnarray}
The recursion formula for the  twisted trigonometrical sum $ T_{2m}^{t}$, may be derived by setting $ l= N/2$ in Eq. (\ref{t28})
\begin{eqnarray}
\label{t29}
T_{2m}^{t}&=&\sum_{n=1}^{N-1}(-1)^{n+1}\frac{1}{\sin^{2m}(n\pi/N)}\nonumber\\&=&\sum_{s=1}^{m-1}(-1)^{s+1 }\frac{N/2}{N/2+s}\binom {N/2+s} {N/2-s}2^{2s}T_{2(m-s)}^{t}\nonumber\\&+&(-1)^{m+1}2^{2m}\frac{(N/2+m-1)!}{(N/2-m)!(2m)!}N/2 \nonumber\\&-&\sum_{n=1}^{N-1}\frac{1}{\sin^{2m}(n\pi/N)}.
\end{eqnarray}
For $m=1$ and $m=2$ the twisted trigonometrical sums are
\begin{equation}
\label{30}
T_{2}^{t}=\frac{N^{2}+2}{6}
\end{equation}
and
\begin{equation}
\label{31}
T_{4}^{t}=\frac{7N^{4}+40N^{2}+88}{360},
\end{equation}
respectively. In obtaining these results we used the expressions for $T_{2}$ and $T_{4}$. These twisted trigonometrical sums appeared  earlier as coefficients of certain generating function  \cite{ Zagier}. Using the recursion formula Eq. (\ref{t29}), the twisted trigonometrical sum $ T_{6}^{t}$ is
\begin{equation}
\label{32}
T_{4}^{t}=\frac{31N^{6}+294N^{4}+1344N^{2}+3056}{15120}. 
\end{equation}
The twisted trigonometrical sum formula given by  Eq. (\ref{t29}), implies that the dimension of the twisted space of the conformal blocks is may be deduced for any genus $g\geq2$, through the following formula 
\begin{eqnarray}
\label{t33}
dimV_{g,k}^{t}&=&\Big(\frac{k+2}{2}\Big)^{g-1}\sum_{n=1}^{k+1}(-1)^{n+1}\frac{1}{\sin^{2g-2}n\pi/(k+2)}\nonumber\\&=&\Big(\frac{k+2}{2}\Big)^{g-1}\sum_{s=1}^{g-2}(-1)^{s+1 }\frac{(k+2)/2}{(k+2)/2+s}\binom {(k+2)/2+s} {(k+2)/2-s}2^{2s}T_{2g-2-2s}^{t}\nonumber\\&+&(-1)^{g}2^{2g-2}\Big(\frac{k+2}{2}\Big)^{g-1}\frac{((k+2)/2+g-2)!}{((k+2)/2-g+1)!(2g-2)!}(k+2)/2 \nonumber\\&-&dimV_{g,k}.
\end{eqnarray}
Note that, the relation  between $dimV_{g,k}^{t} $ and $ dimV_{g,k}$ is expected from the simple identity $$dimV_{g,k-2}^{t}= dimV_{g,k-2}-2^{g}dimV_{g,k/2-1},$$ where $k$ even. The  formula  by Zagier \cite{Zagier}   for $dimV_{g,k-2}$,  may be obtained using the following generating function 
$$ \sum_{g=1}^{\infty}dimV_{g,k-2}\Big(\frac{2}{k}\sin^{2}x\Big)^{g-1}=\frac{k\sin (k-1)x}{\sin kx \cos x }.$$
\section{The corner-to-corner resistance  and the Kirchhoff index of a $ 2\times N $ resistor network}

\ \ In general, it is hard to have a closed-form expression for the two-point resistance of a resistor network, however, if the latter has certain symmetries like circulant resistor network, then this may be possible \cite{Chair1}. The situation gets more and more complicated in two and three dimensional resistor networks \cite{Wu}, as the exact  two-point resistance are expressed in terms of the double and triple summations.  It turns out that the  recently developed techniques by the author \cite{Chair1},  may be used  to obtain exact formula for the two-point resistance of the first non-trivial  $ 2\times N $ resistor network \cite{Chair2}. In this section, we derive an exact formula for the corner-to-corner resistance and the total effective  resistance of a $ 2\times N $ resistor network. The total effective resistance  is also known as the  Kirchhoff index, this index  was introduced in chemistry  as a molecular structure descriptor, it is used  for discriminating among different molecules with similar shapes and structures \cite{Randic}. At the moment, the only   exact two-point resistance not written as a double summation of an $ M\times N $ resistor network is the asymptotic expansion of the corner-to-corner resistance \cite{Essam, Huang}.  It is known that the value of  the asymptotic expansion of the corner-to-corner resistance of a rectangular resistor network  provides a lower bound to the resistance of compact percolation clusters \cite{ Domany}.
\subsection{The exact evaluation of the corner-to-corner resistance }

\ \  The exact expression for the resistance between two nodes of a rectangular network of resistors with
free boundary conditions was given by Wu \cite{Wu}. Suppose that  the resistances in the two spatial directions
are r = s = 1, then the resistance  ${R}_{\,\rm free}$ between two nodes ${\bf r}_1=(x_1, y_1)$
and ${\bf r}_2=(x_2, y_2)$  is 
\bea
&&R_{\{M\times N\}}^{\,\rm free}({\bf r}_1,{\bf r}_2) = \frac {1} {N} \Big| x_1 -x_2 \Big|  + \frac 1 {M} \Big| y_1 - y_2 \Big| +\frac 2 {MN} \nonumber\\
&&\times{\sum_{m=1}^{M-1}\sum_{n=1}^{N-1}
\frac {\Big[\cos\Big(x_1+\frac 1 2\Big)\tm \cos\Big(y_1+\frac 1 2\Big)\tn
 - \cos\Big(x_2+\frac 1 2\Big)\tm \cos\Big(y_2+\frac 1 2\Big)\tn
\Big]^2 } 
{ (1-\cos \tm )  +(1-\cos \tn )  } } ,\nonumber \\
\label{cc1}
\eea
where
\bea
\tm= \frac{m\pi} M, \hskip1cm \tn= \frac{n\pi} N.\nonumber
\eea 
In order to compute the corner-to-corner resistance of a $2 \times N$  resistor network, we set $ M = 2$, ${\bf r}_1=(0, 0)$
and ${\bf r}_2=(1, N-1)$ into Eq. (\ref{cc1}), and so
the double sum of the above equation is reduced to a single sum, and the corner-to-corner
resistance may be written as
\bea
R_{\{2\times N\}}^{\,\rm free}((0,0),(1,N-1))&=&\frac{1}{N}+\frac{N-1}{2} \nonumber\\&+& \frac{1}{3N}\sum_{n=1}^{N-1} \frac{(1+(-1)^{n})(1+\cos n\pi/N)}{2(1-2/3\cos^{2}n\pi/2N)}.
\label{cc2}
\eea
For $N$ even, the sum over  $n$ may be reduced to
\bea
\frac{2}{3N}\sum_{n=1}^{N/2-1} \frac{\cos^{2} n\pi/N}{(1-2/3\cos^{2}n\pi/N)}&=&\frac{1}{3N}\sum_{n=1}^{N-1} \frac{\cos^{2} n\pi/N}{(1-2/3\cos^{2}n\pi/N)}\nonumber\\&=&\frac{1}{3N}\sum_{j=0}^{\infty}(2/3)^{j}\sum_{n=1}^{N-1}\cos^{2(j+1)} n\pi/N,
\label{cc3}
\eea
to evaluate this sum, we follow closely the method developed by the author in \cite{Chair1}.  As explained in \cite{Chair1},
the formula for the sum $\sum_{n=1}^{N-1} \cos^{2J} n\pi/N $,  given by Schwatt \cite{Schwatt} is not the right one to use, we use instead the formula 
\bea
\sum_{n=1}^{N-1} \cos^{2J} n\pi/N =-1+\frac{N}{2^{2j-1}}\sum_{p=1}^{[J/N]}\binom {2J}{J-pN} +\frac{1}{2^{2j}}\binom {2J}{J},
\label{cc4}
\eea
thus, the sum contribution to the corner-to-corner resistance using the residue representation of binomials is
\bea
\frac{1}{3N}\sum_{j=0}^{\infty}(2/3)^{j}\sum_{n=1}^{N-1}\cos^{2(j+1)} n\pi/N&=&-\frac{1}{N}+\sum_{j=0}^{\infty}\hbox{res}_{w}\frac{(1+w)^{2j}}{(6w)^{j}}\frac{w^{N}}{w(1-w^{N})}\nonumber\\&+&\frac{1}{2}\Big[\hbox{res}_w (1-4w){^{-1/2}}\sum_{j=0}^{\infty}(1/6w)^{j}{w^{-1}}-1\Big]\nonumber\\&=&-\frac{1}{N}+\sqrt{3}\frac{(2-\sqrt{3})^{N}}{1-(2-\sqrt{3})^{N}}+\frac{1}{2}(\sqrt{3}-1).
\label{cc5}
\eea
Finally, the corner-to-corner resistance of  $ 2\times N $ resistor network becomes,
\bea
R_{\{2\times N\}}^{\,\rm free}((0,0),(1,N-1))&=&\frac{N-1}{2} +\sqrt{3}\frac{(2-\sqrt{3})^{N}}{1-(2-\sqrt{3})^{N}}+\frac{1}{2}(\sqrt{3}-1).
\label{cc6}
\eea
 It is not difficult to see that this formula is also valid for $N$ odd.
  
  \ Examples. For $N=2,3, 4$ our formula  Eq. (\ref{cc6}), gives
 \bea
 R_{\{2\times 2\}}^{\,\rm free}((0,0),(1,1))&=&1\nonumber\\R_{\{2\times 3\}}^{\,\rm free}((0,0),(1,2))&=&1.4\nonumber\\R_{\{2\times 4\}}^{\,\rm free}((0,0),(1,3))&=&1.875,
 \label{cc7}
 \eea 
these results are in a full agreement with Eq. (\ref{cc2}).
\subsection{The Kirchhoff index}

\ \ The  computation of  the total effective resistance of a  $ 2\times N $ resistor network, that is, the  Kirchhoff index, may be computed in two ways.  It may be evaluated  by summing over all effective resistances between nodes of a given resistor  network, or alternatively by summing over all eigenvalues of a Laplacian associated with resistor  network \cite{Gutman}. So,  we do not have to know the effective resistance between each node to compute the total effective resistance of a resistor network. The formula that gives  the Kirchhoff index of a resistor network in terms of  the eigenvalues is
 $$ Kf(G)=N\sum_{n=1}^{N-1}\frac{1}{\lambda_n},$$  where $\lambda_{n}$ are the eigenvalues of the Laplacian of the network, or the graph $G$ made of nodes and edges considered as  unit resistors.  Our network is given by the cartesian product $ 2\times N$, that is, made of two path lines with $N$ nodes, and $N$ path lines with two nodes. Now,  the Kirchhoff index of a path line is    $$Kf(P_{n})=N\sum_{n=1}^{N-1}\frac{1}{4\sin^2(n\pi/2N)}=\frac{N}{8}\Big[ \sum_{n=1}^{2N-1}\frac{1}{\sin^2(n\pi/2N)}-1\Big]=\frac{N^{3}-N}{6}.$$   
  Thus, the contribution from these path lines is $N+\frac{N^{3}-N}{3}$, by connecting the system together, then the corresponding  eigenvalues of the laplacian are $\lambda_{1,n}= 3(1-2/3\cos^{2}n\pi/2N)$. As a consequence, the Kirchhoff index  of a  $ 2\times N$, resistor network can be written as
\bea 
Kf(2\times N)=N+\frac{N^{3}-N}{3}+N\sum_{n=1}^{N-1}\frac{1}{3(1-2/3\cos^{2}n\pi/2N)},
\label{Kf1}
\eea
Note that, our simple deduction of this expression gives the same value of the Kirchhoff index given by theorem 4.1 in \cite{Yang}. The above sum seems  difficult to  evaluate, however,  using a simple trick, we will be able to get a closed form for the Kirchhoff index. To that end, let us write
\bea
\label{Kf2}
\sum_{n=1}^{N-1}\frac{1}{(1-2/3\cos^{2}n\pi/2N)}&=&\sum_{n=1}^{N/2-1}\frac{1}{(1-2/3\cos^{2}n\pi/N)}\nonumber\\&+&\sum_{n=1}^{N/2}\frac{1}{(1-2/3\cos^{2}(2n-1)\pi/2N)},
\eea
where $N$ is assumed to be even. The first sum may be 
carried out using the following trick;
\bea
\sum_{j=0}^{\infty}(2/3)^{j}\sum_{n=1}^{N-1}\cos^{2(j+1)} n\pi/N&=&\frac{3}{2}\Big[\sum_{j=0}^{\infty}(2/3)^{j}\sum_{n=1}^{N-1}\cos^{2j} n\pi/N-(N-1)\Big].
\label{Kf3}
\eea
Now, the sum on the left-hand side was computed before, see Eq. (\ref{cc5}), then one may deduce
\bea
\sum_{n=1}^{N-1}\frac{1}{(1-2/3\cos^{2}n\pi/N)}&=&\sum_{j=0}^{\infty}(2/3)^{j}\sum_{n=1}^{N-1}\cos^{2j} n\pi/N \nonumber\\&=&-3+\frac{6N(2-\sqrt{3})^{N}}{\sqrt{3}(1-(2-\sqrt{3})^{N})}+\sqrt{3}N
\label{Kf5}.
\eea
and so,
\bea
\sum_{n=1}^{N/2-1}\frac{1}{(1-2/3\cos^{2}n\pi/N)}&=&\frac{1}{2}\Big[\sum_{n=1}^{N-1}\frac{1}{(1-2/3\cos^{2}n\pi/N)}-1\Big]\nonumber\\&=&-2+\frac{3N(2-\sqrt{3})^{N}}{\sqrt{3}(1-(2-\sqrt{3})^{N})}+\frac{\sqrt{3}}{2}N.
\label{Kf6}
\eea
Using the identity \cite{Chair1},
\begin{eqnarray}
\label{Kf7}
\sum_{n=1}^{N/2}\cos^{2j}(2n-1)l\pi/N&=&
\frac{N}{2^{2j+1}}\binom {2j}{j}+\frac{N}{2^{2j}}\sum_{p=1}^{[j/2N]}\binom {2j}{j-2pN}\nonumber\\&-&\frac{N}{2^{2j}}\sum_{p=1}^{[j/2N]}\binom {2j}{j-(2p-1)N},
\end{eqnarray}
 and by following similar steps as in the  above computations, then, one may show
 \bea
 \sum_{n=1}^{N/2}\frac{1}{(1-2/3\cos^{2}(2n-1)\pi/2N)}&=&\frac{3N(2-\sqrt{3})^{2N}}{\sqrt{3}(1-(2-\sqrt{3})^{2N})}-\frac{3N(2-\sqrt{3})^{N}}{\sqrt{3}(1-(2-\sqrt{3})^{2N})}\nonumber\\&+&\frac{\sqrt{3}}{2}N.
 \label{Kf8}
 \eea
 Finally, the exact expression of the Kirchhoff index of a $2\times N$ reeds
\bea
Kf(2\times N)=N+\frac{N^{3}-N}{3}+\frac{N}{3}\Big[-2+\frac{6N(2-\sqrt{3})^{2N}}{\sqrt{3}(1-(2-\sqrt{3})^{2N})}+\sqrt{3}N\Big].
\label{Kf9}
\eea 
One can show that, the above formula for the Kirchhoff formula is valid for $N$ odd as well.

Example, For $N=1,2,3,4,5$, the Kirchhoff indices are respectively, 
 \bea
Kf(2\times 2)&=&5\nonumber\\Kf(2\times 3)&=&14.2 \nonumber\\Kf(2\times 4)&=&30.57142857
\nonumber\\Kf(2\times 5)&=&56.10047847
\eea 
 These results are in complete agreement with those obtained using formula given by Eq. (\ref{Kf1}), or theorem 4.1 of reference \cite{Yang}.
  \section{Some  trigonometrical sums related to number theory}
  
  \ \  In this section other class of  trigonometrical sums  will be evaluated using similar techniques as in the previous sections. Some of these trigonometrical sums are related to number theory. We  will start with the following sum
$$S(l):=\sum_{n=1}^{N-1}(-1)^n\frac{\sin^2(nl\pi/N)}{\sin^2(n\pi/N)}, $$ which is the alternating  sum associated with the sum  $R(l)$ given in Eq. (\ref{t1}). This sum has the following closed formula
\begin{proposition}
\begin{equation}
\label{s}
S(l)=\sum_{n=1}^{N-1}(-1)^n\frac{\sin^2(nl\pi/N)}{\sin^2(n\pi/N)}=-l^2
\end{equation}
\end{proposition}
To derive the above formula, we follow  similar computations carried out for $R(l)$, except that this time the sum over $n$ is  non-vanishing only if $N$ is even
\begin{equation} 
\label{a1}
\sum_{n=1}^{N-1}(-1)^n\sin^{2(s-1)}(n\pi/N)=\frac{1}{2^{2(s-1)-1}}\sum_{t=1}^{s-1}(-1)^{t +1}\binom {2(s-1)}{s-1-t} +\frac{-1}{2^{2(s-1)}}\binom {2(s-1)} {s-1},
\end{equation}
 Comparing Eq. (\ref{a1}) and Eq. (\ref{t3}), and using the previous results, then,  without any further computations,  the formula for the trigonometrical sum $S(l)$ is obtained. 
 Due to the the symmetry enjoyed by $ S(l)$, $S(l)=S(N-l)$, the right hand side of equation (\ref{s}) should be read with this constraint, that is  for  both $l$, and $N-l$,  $ S(l)=-l^2$ . Next, let us consider the sums $ S_{1}(l):=\sum_{n=1}^{N-1}\frac{\sin(nl\pi/N)}{\sin(n\pi/N)}$ and $S_{2}(l):=\sum_{n=1}^{N-1}(-1)^n\frac{\sin(nl\pi/N)}{\sin(n\pi/N)}$ that are closely related. We will prove that  closed formulas for the sums  $ S_{1}(l)$ and $ S_{2}(l)$ are given by
 \begin{theorem}
\begin{equation} 
\label{s1}
S_{1}(l)=\sum_{n=1}^{N-1}\frac{\sin(nl\pi/N)}{\sin(n\pi/N)}=\left\{\begin{array}{cl}N-l & \text{for } l  \text{odd }\\
0 &  \text{for } l  \text{even },
\end{array} \right.
\end{equation}
\begin{eqnarray} 
\label{s2}
S_{2}(l)=\sum_{n=1}^{N-1}(-1)^n\frac{\sin(nl\pi/N)}{\sin(n\pi/N)}=\left\{\begin{array}{cl}-(2l-1) & \text{for } l  \text{odd } \text{and } N \text{even} \\
0 &  \text{for } l  \text{odd } \text{and } N \text{odd}\\
-2l& \text{for } l  \text{even } \text{and } N \text{odd}\\
0& \text{for } l  \text{even } \text{and } N \text{even}.
\end{array} \right.
\end{eqnarray}
\end{theorem}

To prove the first formula, we use the following trigonometrical identity \cite{Hobson},
 \begin{equation} 
\label{Hobson}
\frac{\sin(nl\pi/N)}{\sin(n\pi/N)}=\sum_{s\geq0}(-1)^{s}\binom {l-s-1}{s}2^{l-2s-1}\cos^{l-2s-1}(n\pi/N), 
 \end{equation}
 thus, for $l$ odd, one has
 \begin{eqnarray} 
\label{n1}
S_{1}(l)=\sum_{n=1}^{N-1}\frac{\sin(2l-1)n\pi/N}{\sin n\pi/N}&=&\sum_{s\geq0}(-1)^{s}\binom {2l-2-s}{s}2^{2l-2-2s}\sum_{n=1}^{N-1}\cos^{2l-2-2s}(n\pi/N)\nonumber\\.
\end{eqnarray}
The sum over $n$, may be computed from Schwatt's book \cite{Schwatt}, see Eq. (107), page $221$ to give 
\begin{eqnarray} 
\label{n2}
\sum_{n=1}^{N-1}\cos^{2l-2-2s}(n\pi/N)= \frac{-2}{2^{2l-2-2s}}\sum_{t=1}^{l-1-s}\binom {2l-2-2s}{l-1-s-t} +\frac{N-1}{2^{2l-2-2s}}\binom {2l-2-2s} {l-1-s},
\end{eqnarray}
then, the first contribution to the sum given in Eq. (\ref{n1}) is
\begin{eqnarray} 
\label{n3}
S_{1}(l)^{'}=-2\sum_{s=0}^{l-1}(-1)^{s}\binom {2l-2-s}{s}\sum_{t=1}^{l-1-s}\binom {2l-2-2s}{l-1-s-t}=\nonumber\\-2\hbox{res}_{w=0}\sum_{s=0}^{l-1}(-1)^{s}\binom {2l-2-s}{s}\Big(\frac{1+w}{\sqrt {w}}\Big)^{2l-2-2s}\frac{1}{1-w}\nonumber\\=-2\hbox{res}_{w=0}U_{2l-2}\Big(\frac{1+w}{2\sqrt {w}}\Big)\frac{1}{1-w},
\end{eqnarray}
in obtaining the last line of the above equation we used the expresion for the normalized Chebyshev polynomial of the second kind $U_{n}(\frac{x}{2})= \sum_{k=0}^{[n/2]}(-1)^{k}\binom {n-k}{k}x^{n-2k}$. The residue may be evaluated using $$U_{n}(\frac{x}{2})= \frac{(x+\sqrt{x^2-1})^{n+1}-(x-\sqrt{x^2-1})^{n+1}}{2\sqrt{x^2-1}}, $$  to obtain 
\begin{eqnarray} 
\label{n4}
S_{1}(l)^{'}=-2\hbox{res}_{w=0} \frac{1}{w^{l-1}}\frac{1}{(1-w)^2}=-2(l-1).
\end{eqnarray}
Similarly, the second contribution reads
\begin{eqnarray} 
\label{ns3}
S_{1}(l)^{''}&=&(N-1)\hbox{res}_{w=0}U_{2l-2}\Big(\ \frac{1+w}{2\sqrt {w}}\Big)\frac{1}{w}\nonumber\\&=& N-1.
\end{eqnarray}
Therefore, combining these contributions, the closed formula for $ S_{1}(l)$ is 
\begin{eqnarray} 
\label{n00}
S_{1}(l)=\sum_{n=1}^{N-1}\frac{\sin(2l-1)n\pi/N}{\sin n\pi/N}=N-(2l-1).
\end{eqnarray}
It is not difficult to show that there is no contribution to the sum $ S_{1}(l)$ for $l$ even. In proving the second  formula for $S_{2}(l)$ Eq. (\ref{s2}), one notes that in evaluting the sum over $n$ in $$S_{2}(l)=\sum_{s\geq0}(-1)^{s}\binom {l-s-1}{s}2^{l-2s-1}\sum_{n=1}^{N-1}(-1)^n\cos^{l-2s-1}(n\pi/N),$$  turns out to depend on both $l$, $N$ unlike the previous case. By using formulas  given by Eq (113), and  Eq (114) in \cite{ Schwatt}, we have
\begin{eqnarray} 
\label{n5}
\sum_{n=1}^{N-1}(-1)^n\cos^{l-2s-1}(n\pi/N)&=&\frac{-2}{2^{2l-2-2s}}\sum_{t=1}^{l-1-s}\binom {2l-2-2s}{l-1-s-t} -\frac{1}{2^{2l-2-2s}}\binom {2l-2-2s} {s-l-1},\nonumber\\
\end{eqnarray}
for $l$ odd, $ N$ even, and the sum vanishes for $l$ odd, $ N$ odd. If $l$ is even, then, the above sum is non-vanishing only  for $N$ odd

Therefore, for $l$ odd it follows from  Eq. (\ref{n5}) that we have  
\begin{eqnarray} 
\label{n7} 
S_{2}(l)&=&\sum_{n=1}^{N-1}(-1)^n\frac{\sin (2l-1)n\pi/N}{\sin(n\pi/N)}\nonumber\\&=&-2\hbox{res}_{w=0} \frac{1}{w^{l-1}}\frac{1}{(1-w)^2}-\hbox{res}_{w=0} \frac{1}{w^l}\frac{1}{(1-w)}=-(2l-1),
\end{eqnarray}
 for $l$ even and  $N$ odd, one has the following identity
\begin{eqnarray} 
\label{n6}
\sum_{n=1}^{N-1}(-1)^n\cos^{l-2s-1}(n\pi/N)&=&\frac{-2}{2^{2l-1-2s}}\sum_{t=0}^{l-1-s}\binom {2l-1-2s}{l-1-s-t},
\end{eqnarray}
from which
\begin{eqnarray} 
\label{n0} 
S_{2}(l)=\sum_{n=1}^{N-1}(-1)^n\frac{\sin(2nl\pi/N)}{\sin(n\pi/N)}=-2\hbox{res}_{w=0} \frac{1}{w^{l}}\frac{1}{(1-w)^2}=-2l,
\end{eqnarray}
 these results were recently verified by simulations without proof in connection with number theory  \footnote{Anonymous author  working on characters of a finite field and the Polya-Vinogradov inequality.}. It is interesting to note that the closed formula for $ S_{2}(l)$  may be expected  since if we   let  $l$ go to $ N-l$ in  $ S_{1}(l)$, then, $S_{2}(l)=- l$ . Now, we will consider the following non-trivial and interesting trigonometrical sums, $$ F_{1}(N,l,2):= \sum_{n=1}^{N-1}\frac{\sin(nl\pi/N)}{\sin(n\pi/N)}\frac{\sin(2nl\pi/N)}{\sin(2n\pi/N)},$$ and 
 $$ F_{2}(N,l,2):= \sum_{n=1}^{N-1}(-1)^{n}\frac{\sin(nl\pi/N)}{\sin(n\pi/N)}\frac{\sin(2nl\pi/N)}{\sin(2n\pi/N)},$$
 where $ F_{1}(N,N-l,2)= F_{2}(N,l,2)$, and $ N$ is assumed to be odd. So, if $ F_{2}(N,l,2)$ is known, then,  $ F_{2}(N,l,2)$ may be obtained and vice-versa. In the  rest of this paper, we show that both the trigonometrical sums may be evaluated to give the following closed formulas
\begin{theorem}
\begin{eqnarray} 
\label{f1}
F_{1}(N,l,2)& = &\sum_{n=1}^{N-1}\frac{\sin(nl\pi/N)}{\sin(n\pi/N)}\frac{\sin(2nl\pi/N)}{\sin(2n\pi/N)}\nonumber\\&=&-\frac{1}{2}(3l-2)(3l-3)+\frac{1}{2}(l-1)(l-2)-l+\frac{1}{2}(1-(-1)^l)\nonumber\\&+&\frac{N-1}{2}\Big(2l-1-(-1)^l\Big)+N\Big(3l-2-N+\frac{1}{2}(1-(-1)^{l-N})\Big)\nonumber\\
\end{eqnarray}
where  $l$  is odd  and  the sum vanishes for even l, also, note that the last term namely the coefficient of $N$ is different from zero only if $ 3l-2>N$. The closed formula for $ F_{2}(N,l,2)$ reads
\begin{eqnarray} 
\label{f2}
 F_{2}(N,l,2)&=& \sum_{n=1}^{N-1}(-1)^{n}\frac{\sin(nl\pi/N)}{\sin(n\pi/N)}\frac{\sin(2nl\pi/N)}{\sin(2n\pi/N)}\nonumber\\&=&-\frac{1}{2}3l(3l-1)+\frac{1}{2}l(l-1)-l+N\Big(3l-\frac{(N+1)}{2}+\frac{1}{2}(1-(-1)^{l-\frac{(N+1)}{2}})\Big),\nonumber\\,
\end{eqnarray}
where  $l$  is even  and the sum vanishes for odd l, also, note that the last term whose  coefficient is $N$ is different from zero only if $ 3l>\frac{N+1}{2}$
\end{theorem} 
   To prove the first formula, we note  that $  F_{1}(N,2l,2)=0$, and hence the only sum  to consider is the sum $  F_{1}(N,2l-1,2)$. The latter may be written as 
 \begin{eqnarray} 
\label{n8}
 F_{1}(N,2l-1,2)&=&\sum_{n=1}^{N-1}\frac{\sin(n(2l-1)\pi/N)}{\sin(n\pi/N)}\frac{\sin(2n(2l-1)\pi/N)}{\sin(2n\pi/N)}\nonumber\\&=&\sum_{s,k\geq0}^{l-1}(-1)^{s+k}\binom {2l-2-s}{s}\binom {2l-2-k}{k}2^{2(2l-2)-2(s+k)}\nonumber\\&\times&\sum_{j=0}^{2l-2-2k}(-1)^{j}2^j\binom {2l-2-2k}{k}\sum_{n=1}^{N-1}\cos^{2l-2-2(s-j)}(n\pi/N).
\end{eqnarray}
The sum over $n$, formally looks like that given in Eq. (\ref{n2}), however, the variable $t$, may be a multiple of $N$ and in that case the Schwatt's formula given by Eq (107), does not work it has to be modified slightly. The formula that takes into account this fact may be shown to be given by
\begin{eqnarray} 
\label{n9}
\sum_{n=1}^{N-1}\cos^{2l-2-2(s-j)}(n\pi/N)&= &\frac{-2}{2^{2l-2-2(s-j)}}\sum_{t=1}^{l-1-s}\binom {2l-2-2(s-j)}{l-1-(s-j)-t}\nonumber\\& +&\frac{N-1}{2^{2l-2-2(s-j)}}\binom {2l-2-2(s-j)} {l-1-(s-j)}\nonumber\\&+&\frac{2N}{2^{2l-2-2(s-j)}}\sum_{p=1}^{[l-1-(s-j)/N]}\binom {2l-2-2(s-j)} {l-1-(s-j)-pN},
\end{eqnarray}
where the first two terms in the above formula are those expected from Eq.  (\ref{n2}), while the third term is precisely the correction to the formula Eq.  (\ref{n2}) for  $t$  congruent to $N$. Therefore, there are three contributions to the sum given in Eq. (\ref{n8}), the first of which  reads
\begin{eqnarray} 
\label{n10}
F_{1}^{'}(N,2l-1,2)&=&-2\sum_{s,k\geq0}^{l-1}(-1)^{s+k}\binom {2l-2-s}{s}\binom {2l-2-k}{k}2^{2l-2-2k}\nonumber\\&\times&\sum_{j=0}^{2l-2-2k}(-1)^{j}\frac{1}{2^j}\binom {2l-2-2k}{k}\sum_{t=1}^{l-1-s}\binom {2l-2-2(s-j)}{l-1-(s-j)-t}\nonumber\\&=&-2\hbox{res}_{w=0}U_{2l-2}\Big(\frac{1+w}{2\sqrt {w}}\Big)U_{2l-2}\Big(\frac{1+w^2}{2w}\Big)\frac{1}{1-w}\nonumber\\&=&-2\hbox{res}_{w=0}\Big(\frac{1}{w^{3l-3}}-\frac{1}{w^{l-2}}\Big)\frac{1}{(1-w)^2}\frac{1}{1-w^2}\nonumber\\&=&-\frac{1}{2}(3l-2)(3l-3)+\frac{1}{2}(l-1)(l-2)-l+\frac{1}{2}(1-(-1)^l)
\end{eqnarray}
while the second contribution is
\begin{eqnarray} 
\label{n11}
F_{1}^{''}(N,2l-1,2)&=&(N-1)\sum_{s,k\geq0}^{l-1}(-1)^{s+k}\binom {2l-2-s}{s}\binom {2l-2-k}{k}2^{2l-2-2k}\nonumber\\&\times&\sum_{j=0}^{2l-2-2k}(-1)^{j}\frac{1}{2^j}\binom {2l-2-2k}{k}\binom {2l-2-2(s-j)}{l-1-(s-j)}\nonumber\\&=&(N-1)\hbox{res}_{w=0}U_{2l-2}\Big(\frac{1+w}{2\sqrt {w}}\Big)U_{2l-2}\Big(\frac{1+w^2}{2w}\Big)\frac{1}{w}\nonumber\\&=& (N-1)\hbox{res}_{w=0}\Big(\frac{1}{w^{3l-2}}-\frac{1}{w^{l-1}}\Big)\frac{1}{1-w^2}\frac{1}{1-w}\nonumber\\&=&\frac{N-1}{2}\Big(2l-1-(-1)^l\Big).
\end{eqnarray}
 To obtain the last contribution we write the sum over $p$, in Eq. (\ref{n9}), as $$ \sum_{p=1}^{[l-1-(s-j)/N]}\binom {2l-2-2(s-j)} {l-1-(s-j)-pN}=\hbox{res}_{w=0} \Big( \frac{(1+w)^{2l-2-2(s-j)}w^N}{w^{l-(s-j)}(1-w^{N})}\Big),$$ and using the fact that $l\leq N-1$, then, the third contribution may be computed to give
 \begin{eqnarray} 
\label{n12}
F_{1}^{'''}(N,2l-1,2)&=&2N\sum_{s,k\geq0}^{l-1}(-1)^{s+k}\binom {2l-2-s}{s}\binom {2l-2-k}{k}2^{2l-2-2k}\nonumber\\&\times&\sum_{j=0}^{2l-2-2k}(-1)^{j}\frac{1}{2^j}\binom {2l-2-2k}{k}\hbox{res}_{w=0} \Big( \frac{(1+w)^{2l-2-2(s-j)}w^N}{w^{l-(s-j)}(1-w^{N})}\Big)\nonumber\\&=&2N\hbox{res}_{w=0}\Big(\frac{1}{1-w^N}U_{2l-2}\Big(\frac{1+w}{2\sqrt {w}}\Big)U_{2l-2}\Big(\frac{1+w^2}{2w}\Big)w^{N-1}\Big)\nonumber\\&=&2N\hbox{res}_{w=0}\frac{1}{1-w^N}\frac{1}{w^{3l-2-N}}\frac{1}{(1-w)(1-w^2)}\nonumber\\&=&N\big(3l-2-N+\frac{1}{2}(1-(-1)^{l-N}\big).
\end{eqnarray}
Note that this will contribute only for $3l-2>N $, and as a result the formula for the sum $ F_{1}(N,2l-1,2)$ is
\begin{eqnarray} 
\label{n13}
F_{1}(N,2l-1,2)&=&\sum_{n=1}^{N-1}\frac{\sin(n(2l-1)\pi/N)}{\sin(n\pi/N)}\frac{\sin(2n(2l-1)\pi/N)}{\sin(2n\pi/N)}\nonumber\\&=&-\frac{1}{2}(3l-2)(3l-3)+\frac{1}{2}(l-1)(l-2)-l+\frac{1}{2}(1-(-1)^l)\nonumber\\&+&\frac{N-1}{2}\Big(2l-1-(-1)^l\Big)+N\Big(3l-2-N+\frac{1}{2}(1-(-1)^{l-N})\Big).
\end{eqnarray}
Having obtained a closed formula for the sum  $ F_{1}(N,2l-1,2)$, we now wish  to prove the formula for the   alternating sum  $ F_{2}(N,l,2) $. 
First, we note that for $N$ odd, the sum is non-vanishing only for $l$ is even. Therefore, the formula for  $ F_{2}(N,l,2) $ becomes
\begin{eqnarray} 
\label{n14}
 F_{2}(N,2l,2&)=&\sum_{n=1}^{N-1}(-1)^{n}\frac{\sin((2l)n\pi/N)}{\sin(n\pi/N)}\frac{\sin((2l)2n\pi/N)}{\sin(2n\pi/N)}\nonumber\\&=&\sum_{s,k\geq0}^{[(2l-1)/2]}(-1)^{s+k}\binom {2l-1-s}{s}\binom {2l-1-k}{k}2^{2(2l-1)-2(s+k)}\nonumber\\&\times&\sum_{j=0}^{2l-1-2k}(-1)^{j}2^j\binom {2l-1-2k}{k}\sum_{n=1}^{N-1}(-1)^n\cos^{2l-1-2(s-j)}(n\pi/N).
\end{eqnarray}
The sum over $n$  may carried out using Eq. (114), in \cite{ Schwatt} with the slight modification as explained before, then, it is not difficult to show
\begin{eqnarray} 
\label{n15}
\sum_{n=1}^{N-1}(-1)^n\cos^{2l-1-2(s-j)-1}(n\pi/N)&=&\frac{-2}{2^{2l-1-2(s-j)}}\sum_{t=0}^{l-1-s}\binom {2l-1-2(s-j)}{l-1-(s-j)-t}\nonumber\\&+&\frac{2N}{2^{2l-1-2(s-j)}}\sum_{p\geq1}\binom {2l-1-2(s-j)} {l-1-(s-j)-\frac{(2p-1)N-1}{2}},\nonumber\\
\end{eqnarray}
where the sum over $p$, may be written as
$$\sum_{p\geq1}\binom {2l-1-2(s-j)} {l-1-(s-j)-\frac{(2p-1)N-1}{2}}=\hbox{res}_{w=0} \Big( \frac{(1+w)^{2l-1-2(s-j)}w^{N/2}}{w^{l-(s-j)+1/2}(1-w^{N})}\Big).$$
By using Eq. (\ref{n14}), computations show that the closed formula for the sum $F_{2}(N,2l,2)$ is
\begin{eqnarray} 
\label{n16}
 F_{2}(N,2l,2&)=&\sum_{n=1}^{N-1}(-1)^{n}\frac{\sin((2l)n\pi/N)}{\sin(n\pi/N)}\frac{\sin((2l)2n\pi/N)}{\sin(2n\pi/N)}\nonumber\\&=&-2\hbox{res}_{w=0}\Big(U_{2l-2}\Big(\frac{1+w}{2\sqrt {w}}\Big)U_{2l-2}\Big(\frac{1+w^2}{2w}\Big)\frac{1}{\sqrt{w}(1-w)}\Big)\nonumber\\&+&2N\hbox{res}_{w=0}\frac{1}{1-w^N}\Big(U_{2l-1}\Big(\frac{1+w}{2\sqrt {w}}\Big)U_{2l-1}\Big(\frac{1+w^2}{2w}\Big)\frac{w^{N/2}}{w}\Big)\nonumber\\&=&-2\hbox{res}_{w=0}\Big(\frac{1}{w^{3l-1}}-\frac{1}{w^{l-1}}\Big)\frac{1}{(1-w)^2}\frac{1}{1-w^2}\nonumber\\&+&2N\hbox{res}_{w=0}\frac{1}{1-w^N}\frac{1}{w^{3l-(N+1)/2}}\frac{1}{(1-w)(1-w^2)}\nonumber\\&=&-\frac{1}{2}3l(3l-1)+\frac{1}{2}l(l-1)-l+N\Big(3l-\frac{(N+1)}{2}+\frac{1}{2}(1-(-1)^{l-\frac{(N+1)}{2}})\Big),\nonumber\\
 \end{eqnarray}
 where the last term whose coefficient is $N$, contributes only for $3l> \frac{(N+1)}{2} $. Let us now, check  that the  formulas   $  F_{1}(N,2l-1,2)$, $F_{2}(N,2l,2)$ are consistent with  symmetry   discussed earlier, that is,  $  F_{1}(N,N-l,2)= F_{2}(N,l,2)$, this in turns implies that the  correctness  of the formulas. To do so, we will give some explicit examples, from the expression of $ F_{1}(N,2l-1,2) $ given in Eq. (\ref{n13}), it is clear that the sum should be $N-1$, for $l=1$ and to check this, one has to take into account that when substituting $l=1$ in the formula,  the last term of  Eq. (\ref{n13}) does not contribute. From the symmetry that relates the two sums, we should have $ F_{2}(N,N-1,2)=N-1$. Indeed, this is the case, we simply let $l=\frac{N-1}{2} $ into Eq. (\ref{n16}), this time, however, the last term of this equation does contribute. An explicit computation shows that  $  F_{2}(N,2,2)= -4$, for $N> 3$, and  $  F_{2}(N,2,2)= 2$, for $N= 3$, it is interesting to note that these two cases for $l=1$ are contained in the last term of Eq. (\ref{n15}), since   for $N> 3$, the last term is equal to $0$, and hence $  F_{2}(N,2,2)= -4$, while for $N= 3$, the last term is equal to $6$, that is, our formula gives the right answer. Using the symmetry, we obtain $ F_{1}(N,N-2,2)=-4$, this can be easily checked using our formula given by Eq. (\ref{n13}), and $l=\frac{N-1}{2} $.
 \section{Conclusion}
 
 \ \ To conclude,  in this paper we used our method in \cite{ Chair1}, to give alternative derivations to  closed formulas for trigonometrical sums that appear in one-dimensional lattice,  and in the proof of the  conjecture of F.  R Scott on Permanent of the Cauchy matrix.  A new derivation of certain trigonometrical sum of the perturbative chiral Potts model is given as well as new recursion formulas of certain trigonometrical sums \cite{Mehta}. By  using these recursion formulas, then, one is able  deduce the Verlinde dimension formulas for the untwisted (twisted) space of conformal blocks of  $SU(2)$ ($SO(3)$)WZW. In this paper, we reported closed-form  formulas for the corner-to-corner resistance and the Kirchhoff index of the first non-trivial two-dimensional resistor network, $2\times N$.  We have also, considered other class of trigonometrical sums, some of which appear in  number theory. Here, we followed  similar formalism as in  \cite{ Chair1}, as a consequence  the non-trivial circulant electrical networks (the cycle and complete graphs are not included) are related to non-trivial trigonometrical sums in number theory. For example in\cite{ Chair1}, we had to introduce certain numbers that we called the Bejaia and Pisa numbers with well known properties so that the trigonometrical sums that arise in the computation of the two-point resistance are written in terms of these numbers nicely. By using the  well known connection between the electrical networks and  the random walks \cite{Doyle}, one may hope to give  interpretations  to some of the trigonometrical sums in number theory other than those associated with the two-point resistance of a given electrical network, since the latter  provides an alternative way to compute the basic quantity relevant to random walks known as  the first passage time,  the expected time to hit a target node for the first time  for a walker starting from a source node \cite{Tetali}.
 \vspace{7mm}

{\bf Acknowledgment:}

I would like to thank Professor Bruce Berndt for reading and making comments on the manuscript. Also, I would like to thank the Abdus Salam centre for Theoretical Physics for supports and
hospitality throughout these years 
\newpage
\bibliographystyle{phaip}

\end{document}